\documentclass[12pt,preprint]{emulateapj}

\usepackage{amsmath}

\begin{document}

\title{Population Synthesis of Black Hole Binaries with Normal-Star Companions: I. Detached Systems 
}
\author{
Yong Shao$^{1,2}$ and Xiang-Dong Li$^{1,2}$}

\affil{$^{1}$Department of Astronomy, Nanjing University, Nanjing 210046, People's Republic of China; shaoyong@nju.edu.cn}

\affil{$^{2}$Key laboratory of Modern Astronomy and Astrophysics (Nanjing University), Ministry of
Education, Nanjing 210046, People's Republic of China; lixd@nju.edu.cn}

\begin{abstract}

Optical observations of normal-stars in binary systems with massive unseen objects
have been proposed to search for candidate black holes (BHs) and provide a direct measurement of their 
dynamical masses.  In this paper, we have performed binary population synthesis calculations to simulate 
the potential population of detached binaries containing BHs and normal-star companions in the 
Galaxy. We focus on the influence of the BH progenitors. In the traditional model, BHs in binaries 
evolve from stars more massive than $ \sim25M_{\odot} $. However, it is difficult for this model 
to produce BH low-mass X-ray binaries. Recent investigations on massive star evolution suggest 
that the BH progenitors may have masses as low as $ \sim15M_{\odot} $. Based on this model, we
provide the expected distributions of various parameters for detached BH binaries with 
normal-star companions, including the component masses, the orbital parameters of the binary systems,
the radial velocity semi-amplitudes, and the astrometric signatures of the optical companions. 
Our calculations show that there are more than thousands of such detached binaries in the Galaxy, 
and hundreds of them are potentially observable systems with luminous 
companions brighter than 20 mag. In addition, detached BH binaries 
are dominated by those with main-sequence companions and only a few percent of them
are expected to have giant companions.

\end{abstract}

\keywords{binaries: general -- stars: black holes --  stars: evolution}

\section{Introduction}

It is believed that there are hundreds of millions of stellar-mass black holes (BHs) in the Galaxy 
\citep{v92,bb94,tww96}. More than forty years has passed since the discovery of the first BH in 
Cygnus X-1 \citep{b72,wm72}. To date, however, only two dozen of BHs have been dynamically 
confirmed. The majority of them are discovered in X-ray binaries \citep{rm06,cj14}, in which the
BH is accreting material from its companion star and emitting X-ray radiation. According to binary
evolution theories, quite a number of binary systems are likely to host a quiescent BH orbiting its 
normal-star\footnote{In this work, a normal-star specifically refers to a star staying at the 
main-sequence or (super)giant stage. } companion,  prior to the X-ray binary phase, since the 
X-ray radiation due to BH accretion is too weak to be detected.  

A promising approach based on radial velocity searches has been proposed to discover BHs in binary 
systems for decades \citep{gz66,tt69}. Until recently, dynamical searches of optical companions are 
broadly used to identify BHs in binary systems. In the globular cluster NGC 3201, \citet{gd18} found a 
main-sequence (MS) turn-off  star orbiting an unseen component with large radial velocity
variations. A deep analysis of the orbital parameters indicated that the unseen object is a potential BH 
with a minimum mass of $ \gtrsim 4M_{\odot} $. 
Based on the spectroscopic and photometric study of the binary AS 386, \citet{km18}
revealed that the binary system has a circular orbit with a period of $ \sim131 $ days,
and contains a B-type star of mass $ 7\pm1M_{\odot} $. Based on the absence of any traces of the secondary 
component, whose mass is larger than $7 M_{\odot} $, \citet{km18} suggested that it is most likely a BH.
By combining radial velocity measurement with photometric variability data, \citet{tk19} discovered 
a candidate BH orbiting a $ \sim 3M_{\odot} $ giant star with an orbital period of $ \sim 83 $ days.
Since BHs discovered in such detached binaries are not subject 
to the effect of possible mass accretion, a large sample in the future can give a clear clue of 
the BH mass function and test the theories of binary evolution and the supernova (SN) mechanism. 

According to some related surveys, the prospect of hunting BHs in binaries with 
visible normal-stars has already been investigated.
For the astrometric satellite \textit{Gaia} over its five year mission, several groups \citep{bcl17,ml17,yk18,yb18} 
predicted that dozens or thousands of binaries with a BH component may be discovered.  Using the photometric 
data of \textit{Transiting Exoplanet Survey Satellite}, \citet{mh18} discussed the potential of identifying BHs
with normal-star companions on tight but detached orbits. Based on spectroscopic observations of \textit{Large sky 
Area Multi-Object fiber Spectroscopic Telescope}, \citet{gm19} proposed a method to search for stellar-mass BH 
candidates in binaries with giant companions.  

Stellar evolution predicts that BHs originate from massive stars with masses $ \gtrsim 20-25 M_{\odot} $ 
\citep{ww95,f12}. The standard formation scenario for BH binaries with initially a low-mass companion 
usually involves a common envelope (CE) phase \citep[see a review by][]{i13} in which the spiral-in 
of the low-mass secondary causes the ejection of the envelope of the primary (BH's progenitor). 
However, it was pointed out that the primary's envelope is too massive for a low-mass secondary to 
strip off \citep{prh03}, probably resulting in the binary merge. In view of the CE phase, 
compact BH binaries with low-mass companions can be formed only if adopting 
high values for the binding energy parameter of the primary envelope \citep{prh03} 
or for the CE ejection efficiency \citep{kh06,yl08}. Another solution to this problem is adopting relatively small masses for 
the BH progenitors \citep{wjl16a}. According to the argument of 
\citet{k14}, stars with initial masses $ \gtrsim 17 M_{\odot} $ may experience
failed explosions and evolve to BHs. \citet{wjl16a} showed that short orbital-period BH low-mass X-ray binaries can be
effectively produced through the standard CE scenario if most BHs are born in failed SNe. 
It has been long known that the pre-SN core structure of a massive star greatly determines the final fate of
either explosion or implosion \citep{bhf95}. Recently, some numerical simulations 
\citep[e.g.,][]{oo11,uj12,pt15,ej16,se16} showed that the landscape of neutrino-driven explosions is 
strongly manipulated by the final core structure of massive stars, and there is no clean threshold to separate
the outcomes of either neutron stars (NSs) or BHs.  Stars with masses $ \sim 15-21M_{\odot} $ still 
have a possibility to eventually implode to a BH; on the contrary, a star massive than $ 20M_{\odot} $ may successfully 
explode to become an NS \citep{rs18}. 
  
In this work, we perform binary population synthesis (BPS) calculations to simulate the Galactic population
of BH binaries with normal-star companions. 
The main goal of this work is to estimate the number and parameter distribution of BH binaries 
that can be potentially discovered through optical observations of the normal-star companions. 
We here only consider the detached systems without the occurrence of Roche lobe overflow. 
In a forthcoming study, we 
will discuss the population of mass-transferring BH systems (i.e., X-ray binaries) in the Galaxy. 
The remainder of this paper is organized as follows.  In Section 2, we introduce the BPS
method, according to which we can generate a large number of BH binaries with normal-star companions. 
We present the calculated results and discussions in Section 3. Finally we conclude in Section~4.

\section{Method}

The X-ray emission in detached systems should be very weak and even undetected due to little 
accretion onto the BHs. The BH masses in such binaries, however, can still be measured
from the motions of the optical companions. 
In order to obtain the potential population of detached BH binaries with a normal-star companion in
the Galaxy, we employ the population synthesis code \textit{BSE} originally developed 
by \citet{h02}. With \textit{BSE} we can simulate the evolution of millions of binary stars with 
different initial parameters. The binary evolution is assumed to start from primordial 
binaries with two zero-age MS stars, and the subsequent evolution will be subject to many physical
processes, e.g. mass and angular momentum transfer, CE evolution, BH formation and natal kicks.
A detail modification of the code has been made by \citet{sl14}, here we notice some important points 
in the following.

We only consider the detached BH binaries formed through the evolution 
of isolated binaries, that is, systems formed through dynamical interactions in global clusters
are not included. During the evolution of the primordial binaries, the primary star first evolves 
to expand, then transfers mass to the secondary. Thus one first needs to determine whether 
the mass transfer is dynamically stable. This is critically dependent on the mass ratio of the 
primary and secondary stars and the mass accretion efficiency of the secondary star.
\citet{sl14} built three mass transfer modes to deal with the mass exchange between binary components, 
among which the rotation-dependent mode (assuming the accretion efficiency of the secondary to 
be dependent on its rotating velocity) appears to better reproduce the observed
parameter distribution of Galactic binaries including BH$ -$Be star systems \citep{sl14},
Wolf-Rayet star$  -$O-type star systems \citep{sl16}  and NS$-$NS systems \citep{sl18}.
Thus we adopt the rotation-dependent mass transfer mode, in which the accretion 
efficiency of the secondary can be as low as $ < 0.2 $ and the corresponding mass ratio of the primary to 
the secondary for stable mass transfer can reach $ \sim 6 $ \citep{sl14}. Compared with the traditional 
conservative mass transfer, this allows a much larger parameter space for stable mass transfer 
in the primordial binaries.

The binaries experiencing dynamically unstable mass transfer will go into CE evolution. 
The orbital energy of the embedded binary is used to eject the envelope. We adopt the 
standard energy conservation equation \citep{w84} to deal with the orbital decay, taking 
the binding energy parameter $ \lambda $  calculated by \citet{xl10}, and the CE ejection efficiency 
$ \alpha_{\rm CE} $ is assumed to be unity\footnote{Recently \citet{far19}
simulated the inspiral of a $ 1.4M_\odot $ NS inside the envelope of a $ 12M_\odot $ red supergiant star and 
suggested a very high $ \alpha_{\rm CE} $-equivalent efficiency of $ \approx 5 $. Also, \citet{mg18} showed that 
the cosmic merger rates predicted for NS$-$NS binaries are consistent with LIGO-Virgo estimations when adopting
$ \alpha_{\rm CE} = 5$ in population synthesis models. If the $ \alpha_{\rm CE} $ is indeed higher than
unity, it could help resolve the formation problem of  BH low-mass X-ray binaries \citep{prh03,kh06,yl08}.}.
After CE evolution, the remnant binaries may survive
if they do not merge during the spiral-in stages. Stellar mass loss rates of \citet{h00} are employed, 
except for hot OB stars, for which we apply the simulated rates of \citet{v01}.  When the entire envelope
of the primary star is stripped due to binary interactions, we reduce the mass loss prescription of \citet{h95} 
by a factor of 2 for helium stars \citep{kh06}.

At their formation, the BHs may be imparted a natal kick, resulting in eccentric orbits or even
disruption of the binary systems.  The SN processes play a vital role in determining both the magnitude of
the kick velocities and the weight of the BH masses. 
In order to account for the $ \sim 2-5M_{\odot} $ gap between NS and BH masses, \citet{f12} proposed the 
rapid SN mechanism\footnote{Note that the other method of the delayed SN mechanism
in \citet{f12} allows the formation of low mass BHs, so is not included in our calculations.}  assuming that the final mass of a compact object
is contributed by the proto-compact object and the fallback material. With this mechanism stars with masses 
$ \gtrsim 20M_{\odot} $ could evolve to BHs. 
More recently, \citet{se16} suggested that stars with masses as low as $ 15M_{\odot} $ have a chance to implode to be a BH,  
and the average likelihoods for stars with initial masses of $ 15-40M_{\odot} $ and $ 45-120 M_{\odot}$ were 
respectively 0.574 and 0.656 \citep{rs18}. In this case the BH masses were directly obtained as the remnant 
masses of pre-SN stars, or the combined masses of the helium core and a small fraction ($ \lesssim 0. 1$) of the envelope 
if exists.  
In our simulations we consider both prescriptions to deal with the BH masses. The first one is  
$ M_{\rm BH}  = 0.9\, (M_{\rm proto} + M_{\rm fb})$\footnote{The value of 0.9 denotes an efficiency to convert
baryonic to gravitational masses \citep[e.g.,][]{tww96}.}, where  $M_{\rm proto}$ and $M_{\rm fb} $ are
respectively the masses of the proto-compact object and the fallback material. The other one is  
$ M_{\rm BH}  = 0.9\, M_{\rm rem}$, where $M_{\rm rem}$ is the pre-SN remnant masses.
Note that the pre-SN primaries in our BPS calculations are always helium stars without 
any hydrogen envelope due to binary interactions, and the stars with  $M_{\rm rem}= 5M_{\odot} $ 
have initial masses close to $15M_{\odot} $. 
Following the suggestion of \citet{rs18}, we assume that all pre-SN primaries with masses larger than 
$ 5M_{\odot} $ have a probability of 0.6 to form BHs.

For BH natal kicks in the former situation, we use the NS kick velocity
reduced by a factor of $ (1-f_{\rm fb})$, where $ f_{\rm fb} $ is the fallback material fraction (denoted as Model~A). 
Here the kick velocity for NSs is assumed to follow a Maxwellian distribution with a dispersion of 
$ \sigma_{\rm k} = 265\,\rm km\,s^{-1} $ \citep{h05}. In the later situation,
the BH formation does not involve the fallback process, we explore three variations of the BH natal kicks: 
(1) using the NS kick velocity reduced by a factor of $ (3M_{\odot}/M_{\rm BH}) $ (denoted as Model B)\footnote{
In this model, the kick velocities are assumed to be inversely proportional to the BH masses, 
and the minimum BH mass is set to be $ 3M_{\odot} $ \citep{wjl16a}.}; 
(2) obeying a Maxwellian distribution with a dispersion of  $\sigma_{\rm k} = 150\,\rm km\,s^{-1} $
(denoted as Model C); (3) adopting a Maxwellian distribution with a smaller 
dispersion of  $\sigma_{\rm k} = 50\,\rm km\,s^{-1} $ (denoted as Model D). 

\begin{figure*}[hbtp]
\centering
\includegraphics[width=0.7\textwidth]{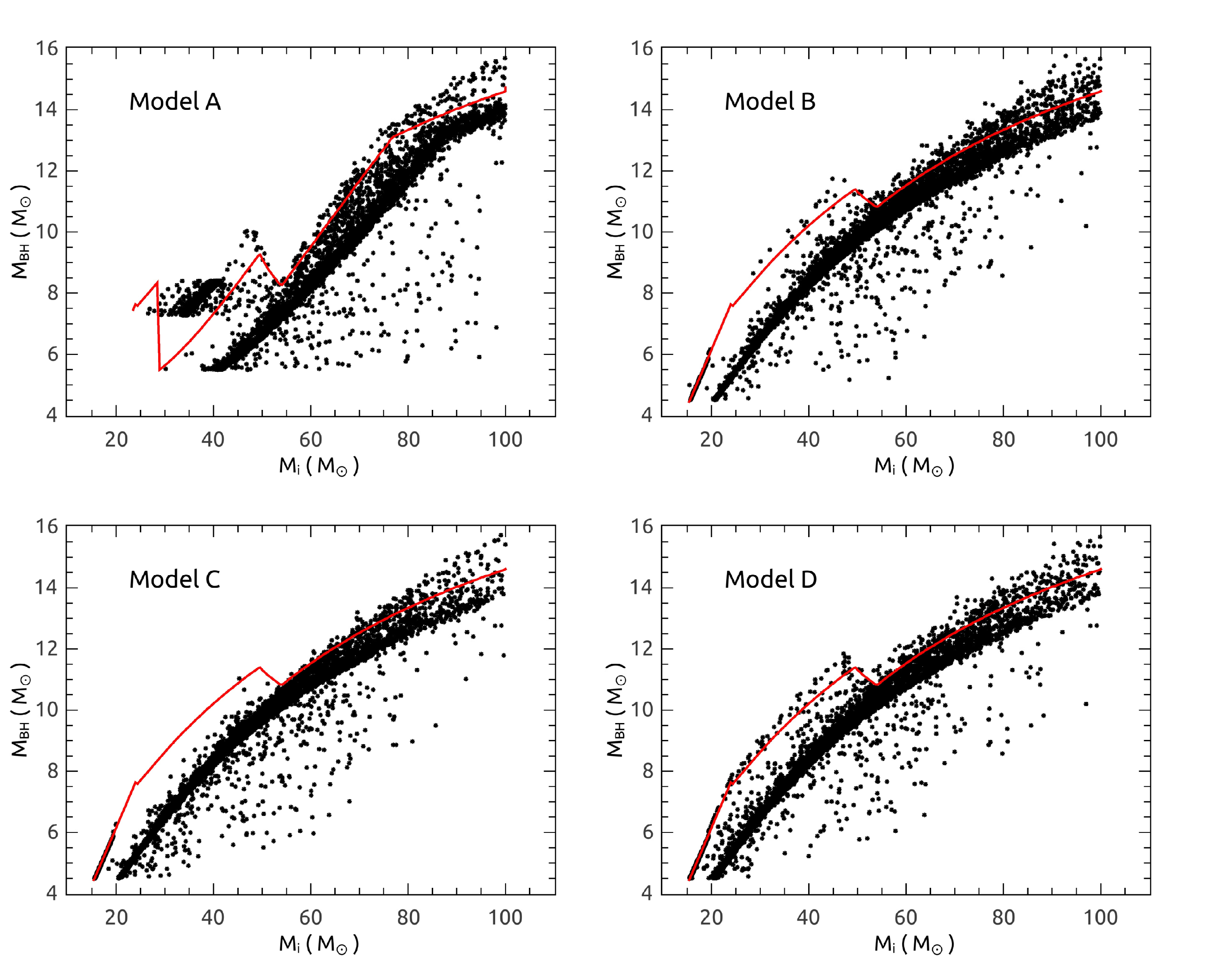}
\caption{Distributions of BH mass $ M_{\rm BH} $ as a function of the initial mass $ M_{\rm i} $ 
of the primary star. The four panels correspond to Models A$-$D. In each panel, the black dots
(with a number of $ 5\times 10^3 $) represent the output of the BPS calculations, and the red curve 
corresponds to the BH mass spectrum for single star evolution.
   \label{figure1}}

\end{figure*}

\begin{table*}
\begin{center}
\caption{Physical inputs of different models, calculated birthrates of incipient BH binaries, and estimated
numbers of detached BH binaries with MS or giant companions in the Galaxy.
\label{tbl-1}}
\begin{tabular}{cccccclllllll}
\\
\hline
Models     & $ M_{\rm BH} $ & $V_{\rm k}$ & $R_{\rm birth} ({\rm yr^{-1}} )$ & $ N_{\rm BH-MS} ^{*}$ &  $ N_{\rm BH-G} ^{*}$ \\
\hline
A   &  $ M_{\rm proto} +  M_{\rm fb} $      & $ \propto1-f_{\rm fb} $   &$9.0\times10^{-5}$ & 470 (260) & 2.4 (1.7) \\
B     & $ M_{\rm rem} $                & $ \propto 3.0/M_{\rm BH} $  & $6.0\times10^{-5}$& 4100 (340) & 160 (11) \\
C     &$ M_{\rm rem} $       &  $ \sigma_{\rm k} = 150\,\rm km\,s^{-1} $    & $4.5\times10^{-5}$ & 4300 (285) & 175 (12) \\
D     &  $ M_{\rm rem} $     & $ \sigma_{\rm k} = 50\,\rm km\,s^{-1} $ &  $1.3\times10^{-4}$ & 12000 (926)  & 595 (48) \\
\hline
\end{tabular}
\end{center}
$^{*}$ The numbers of detached BH binaries with normal-star companions brighter than 20 mag 
are given in parenthesis. \\
\\
\end{table*}

\begin{figure*}[hbtp]
\centering
\includegraphics[width=0.7\textwidth]{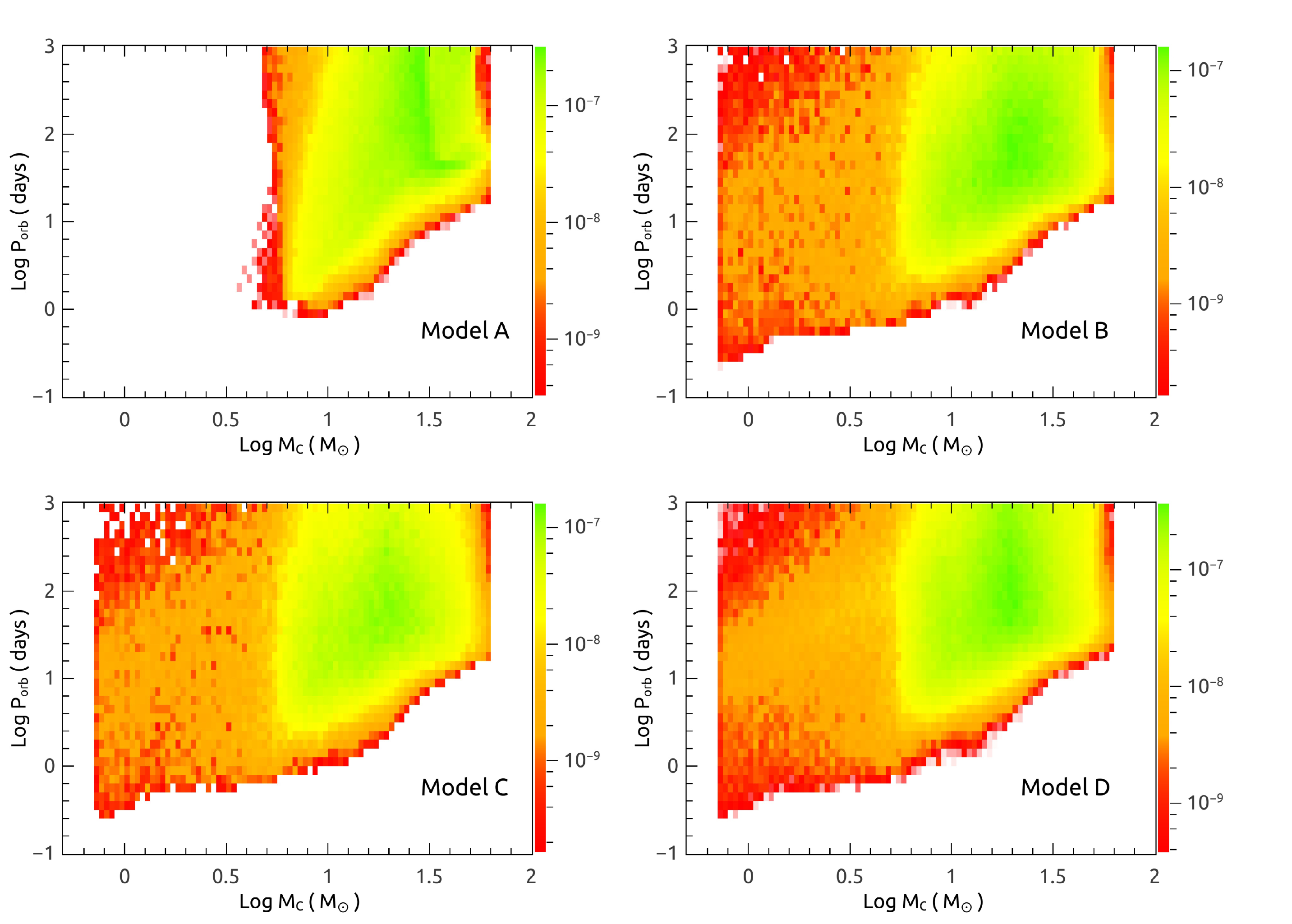}
\caption{The birthrate distributions of incipient BH binaries in the companion mass ($ M_{\rm c} $) vs.
orbital period ($ P_{\rm orb} $) plane, by assuming a constant star formation rate of the Galaxy to be 
$ 3 M_{\odot}\,\rm yr^{-1} $. The four panels correspond to different assumed models, 
the colors in each pixel are scaled according to the weight of the corresponding 
birthrates.   \label{figure1}}

\end{figure*}

\begin{figure*}[hbtp]
\centering
\includegraphics[width=0.7\textwidth]{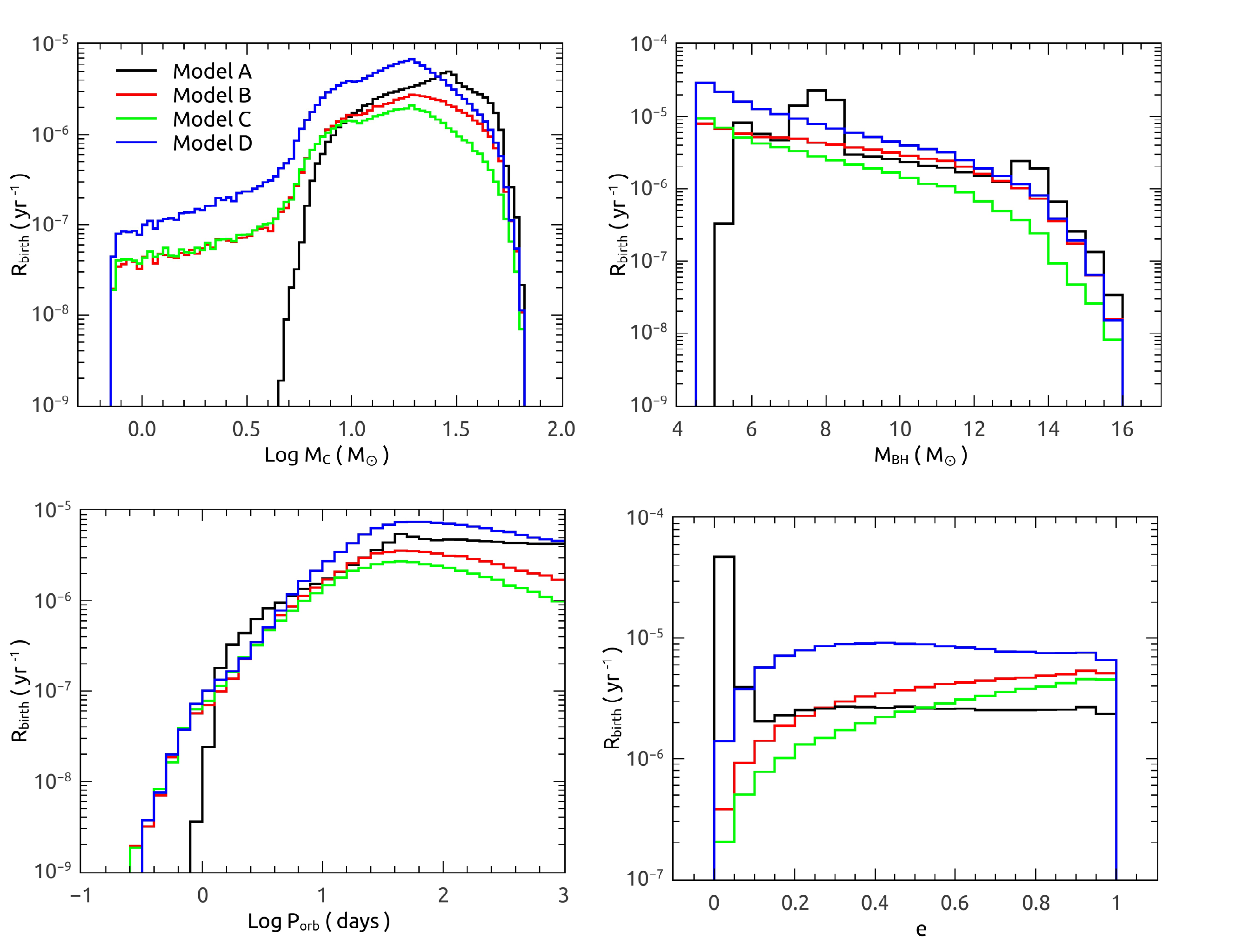}
\caption{The birthrate distributions of incipient BH binaries in the Galaxy, as a function of 
companion mass $ M_{\rm c} $, BH mass $ M_{\rm BH} $, orbital period $ P_{\rm orb} $
and eccentricity $ e $. The colored curves correspond
to different assumed models. 
   \label{figure1}}

\end{figure*}

We simulate the evolution of the primordial binaries by setting the initial parameters as follows. 
For the primary stars, we use the initial mass function (IMF) given by \citet{k93}. For the secondary
stars, we assume a flat mass ratio distribution between 0 and 1.
The distribution of initial orbital separations is assumed to be logarithmically flat \citep{a83}. 
We set the initial orbits of all binaries to be circular, as the outcome of the interactions of 
systems with the same semilatus rectum is almost independent of eccentricity \citep{h02}. 
We assume all stars are initially in binaries.  The primordial binaries are thought to follow 
the distribution of stars in the Galaxy and the BH binaries are simply assumed to be close 
to their birth locations without considering possible motions. The initial metallicity of stars 
is set to be 0.02. Considering the star formation history of the Galaxy, we adopt a constant 
star formation rate of $ 3 M_{\odot}\,\rm yr^{-1} $ over the past 10 Gyr period \citep{sb78,dh06,rw10}. 

If a primordial  binary evolves through a phase that is identified to be a detached BH system,
then such a binary makes a contribution to the birthrate of the specific type of detached BH system.
We follow the method of \citet{h02} to calculate the birthrate for each type of detached BH binaries generated by 
our BPS calculations, which depends on the star formation rate of the Galaxy and the initial parameters
of the primordial binaries.

\section{Results and discussions}

\subsection{Mass spectrum of newborn BHs}

In Figure~1 we show the distributions of BH mass $ M_{\rm BH} $ as a function of the initial mass $ M_{\rm i} $
of the primary star. The four panels correspond to the adopted models A$ - $D. In each panel, the black dots
(with a number of $ 5\times 10^3 $) represent the output of the BPS calculations, and the red curve 
the BH mass spectrum for single star evolution. 
In all the four panels, there is a dip in the red curve around $ 50M_\odot $, which is caused by the very 
strong winds from luminous blue variables. For stars more massive than $ 50M_\odot $, the BH mass 
spectra in both single and binary evolution cases are broadly consistent with each other; 
for stars less massive than $ 50M_\odot $, the BHs evolved from single stars are generally more massive 
than those from binary evolution, since the mass transfer in the primordial binaries decreases the mass of 
the primary stars. However, in Model A, there is an abrupt step at $ 23 M_\odot \lesssim M_{\rm i} \lesssim 28 M_\odot$
in the case of single star evolution and at $ 25 M_\odot \lesssim M_{\rm i} \lesssim 40 M_\odot$ in 
the case of binary evolution, corresponding to the process of direct collapse in the rapid SN mechanism  
(\citealp{f12}; See also \citealp{smb15}).
We obtain $ M_{\rm BH} \sim 5-16M_\odot $ in all the models. \citet{smb15} used the population synthesis code \textit{SEVN} to track the stellar evolution and 
obtained the mass spectrum of stellar-mass BHs.  
Applying the rapid SN model, \citet{smb15} derived the masses of  BHs in the range of 
$ \sim 6-25M_\odot $. This difference may originate from the different treatment of stellar winds.  

\subsection{Formation of incipient BH binaries}

An incipient BH binary is a binary system just after the BH formation. Usually the  
companion is an MS star. Based on the BPS calculations, we create several $ 10^{5} $
incipient BH binaries in each model, then pick out the systems with orbital periods less than $ 10^{3} $ days. 
Figure~2 shows the calculated birthrate distributions of the incipient BH binaries in the 
companion mass $ M_{\rm c} $ vs. orbital period $ P_{\rm orb} $ plane in different models. 
The colors in each pixel are scaled according to the weight of the corresponding 
birthrates. It is clearly seen that the incipient BH binaries with low-mass ($ \lesssim 5M_{\odot} $)
companions are hardly produced in Model A. The initial masses of the BH progenitors in this model are 
larger than $ \sim25M_{\odot} $, so low-mass secondaries are not able to eject the primary's envelopes 
during CE evolution \citep[see also][]{prh03,wjl16a}, leading the binary 
system to merge to be a single star. Almost all of the incipient BH systems produced in Model A 
are the descendants of the primordial binaries experiencing stable mass transfer during the evolution. 
The companion masses distribute in the range of $ \sim 5- 50 M_{\odot} $ and the orbital periods of $\sim1-1000 $ days, 
the overall birthrate in the Galaxy is about $9.0\times10^{-5}\rm\,yr^{-1}$.  In Model B, the  
minimum companion mass is  $ \sim 0.7M_{\odot} $, and  the total 
birthrate of incipient BH systems is $\sim6.0\times10^{-5}\rm\,yr^{-1}$. 
This is because the initial masses of the BH progenitors in this model decrease to $ \sim 15M_{\odot} $ \citep{rs18}, 
so binaries with low-mass secondaries are able to survive CE evolution. When applying other 
kick velocity distributions to the newborn BHs in Models C and D, the distributions of incipient BH binaries 
are similar in the $ M_{\rm c} - P_{\rm orb} $ plane as in Model B, with
the birthrates of   $\sim4.5\times10^{-5}\rm\,yr^{-1}$ and 
$\sim1.3\times10^{-4}\rm\,yr^{-1}$, respectively. 

In Figure~3 we plot the obtained birthrates of the incipient BH binaries as a function of the companion mass, 
BH mass, orbital period and eccentricity in Models A$-$D.  Besides no low-mass companions,
Model A predicts that the BH masses have a peak distribution at $ \sim 7-8M_{\odot} $, and 
the binary systems tend to have nearly circular orbits. The reason is that most of BHs are formed through 
direct collapse without a natal kick. In the other three models, the incipient BH binaries
unsurprisingly have similar parameter distributions. Note that the binaries containing lighter BHs tend to
have higher birthrates due to the IMF. In the following, we will simulate the Galactic population of detached BH
systems with normal-star companions, by tracking the evolution of the incipient BH
binaries until the companions fill their corresponding Roche lobes. For clarification, we only show the 
Galactic population of detached BH binaries in Models A and B, the calculated results for all models 
are summarised in Table~1.

\subsection{Overall population of  detached BH binaries}

A detached BH binary can be discovered through observation of the optical companion. Based on the 
BPS calculations, we can predict the distributions of some observational parameters 
including the apparent magnitude $ m_{\rm V} $, the radial velocity semi-amplitude $ K $ and 
the astrometric signature $ \alpha $. 
We use the stellar mass, surface luminosity and effective temperature of the optical companions
to yield the absolute magnitude $ M_{\rm V} $ and the stellar color $ B-V $.  
By taking into account the interstellar extinction 
$ A_{\rm V} $ in the V-band, the apparent magnitude $ m_{\rm V} $ is given as
\begin{equation}
m_{\rm V} = M_{\rm V} + 5(2+\log D_{\rm kpc}) + A_{\rm V}(D_{\rm kpc}), 
\end{equation}
where $ D_{\rm kpc} $ is the distance of the binary from the Sun normalized by 1 kpc
and we assume $A_{\rm V}(D_{\rm kpc}) = D_{\rm kpc}$
\citep[see also][]{yk18,yb18}.  
The radial velocity semi-amplitude $ K $ of the optical companions is 
\begin{equation}
K  = \sqrt{\frac{G}{(1-e^{2})}} (M_{\rm c}+M_{\rm BH})^{-1/2}a^{-1/2}M_{\rm c}\sin i, 
\end{equation}
where $ G $ is the gravitational constant, $ e $ the eccentricity, 
$ a $ the semi-major axis and $ i $ the orbital inclination of detached BH binaries 
with respect to the Sun. The orientation of binary systems is assumed to have a random distribution. 
The astrometric signature $ \alpha $ is given by 
\begin{equation}
\alpha  = \frac{a_{\rm project}}{D},
\end{equation}
where $ D=D_{\rm kpc} \times 1 \,\rm kpc $, and $ a_{\rm project} $ denotes the projected 
semi-major axis of the companion's orbits \citep[see also][]{bcl17,ml17}.  

\begin{figure}[hbtp]
\centering
\includegraphics[width=0.45\textwidth]{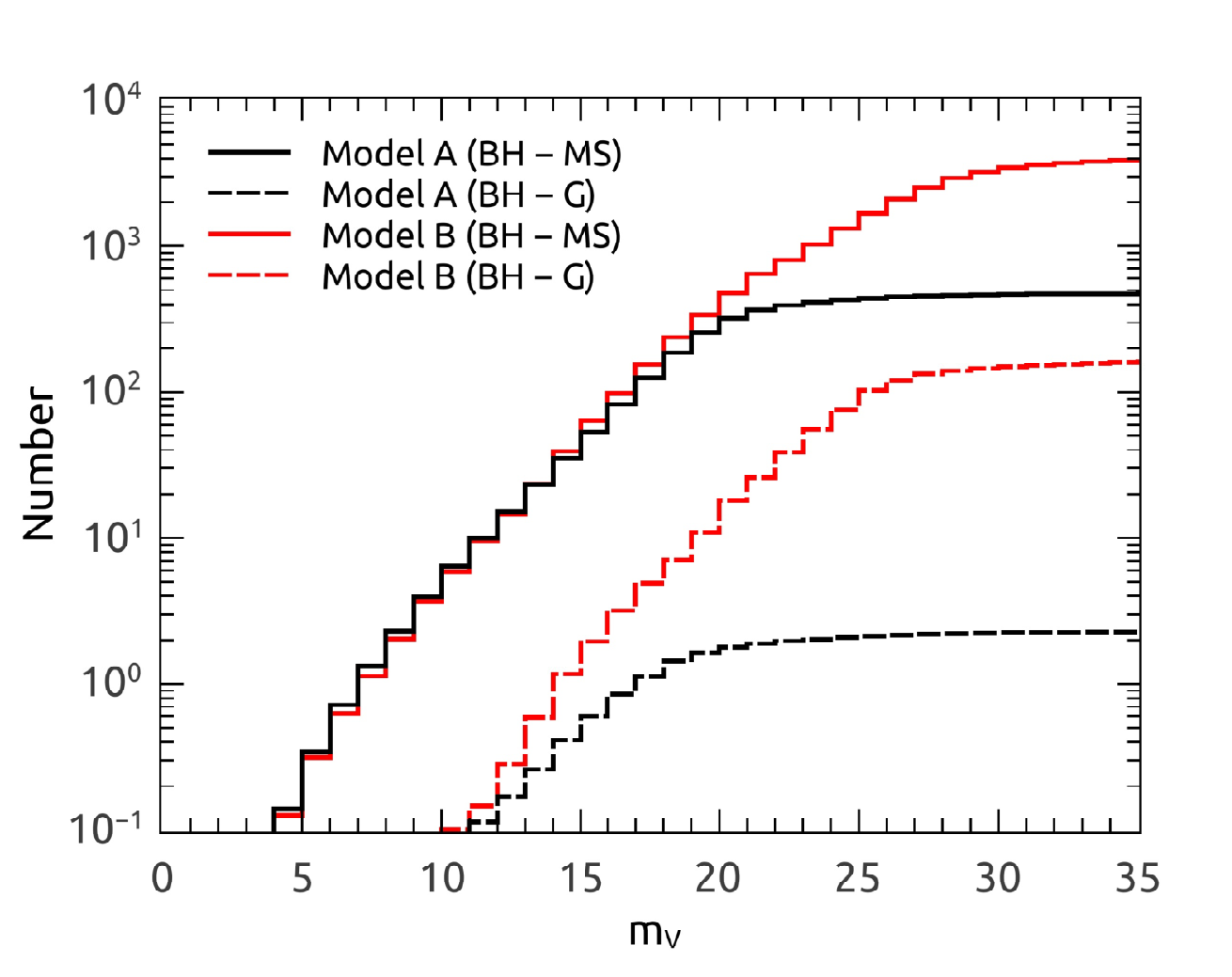}
\caption{Accumulated number distributions for detached BH binaries with normal-star companions in 
the Galaxy as a function of apparent magnitude $ m_{\rm V} $, when gradually 
increasing the $ m_{\rm V} $  by an interval of 1 mag. The black and red curves correspond to Models A
and B, and the solid and dashed curves denote the systems with MS and giant companions, respectively.
   \label{figure1}}

\end{figure}

\begin{figure*}[hbtp]
\centering
\includegraphics[width=1.0\textwidth]{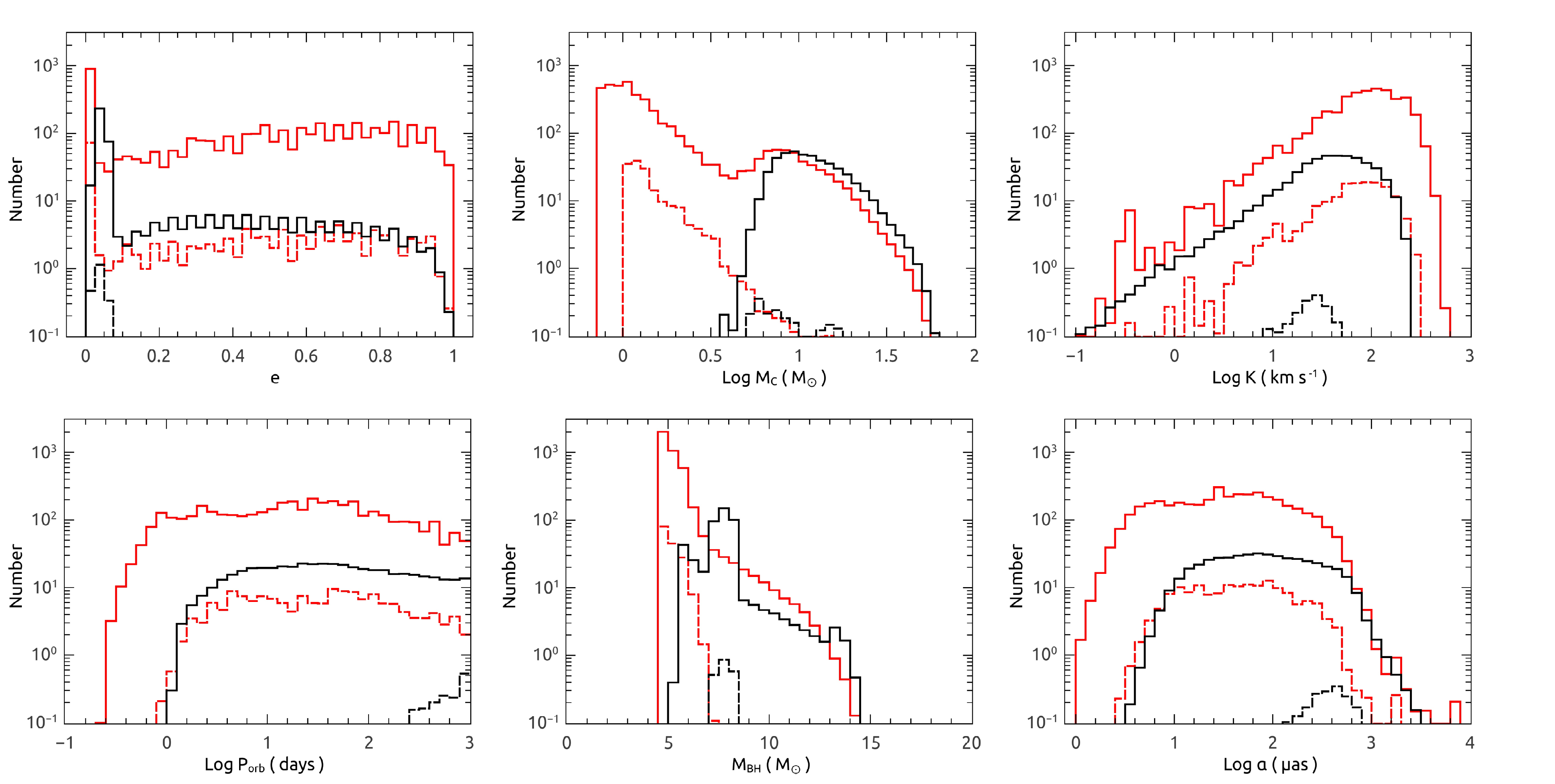}
\caption{Expected number distributions of all detached BH binaries in the Galaxy,
as a function of orbital period $ P_{\rm orb} $, eccentricity $ e $,  companion mass $ M_{\rm c} $, 
BH mass $ M_{\rm BH} $, radial velocity semi-amplitude $ K $ and astrometric signature $ \alpha $. 
The black and red curves correspond to Models A
and B, and the solid and dashed curves denote the systems with MS and giant companions, respectively.
   \label{figure1}}

\end{figure*}

\begin{figure*}[hbtp]
\centering
\includegraphics[width=1.0\textwidth]{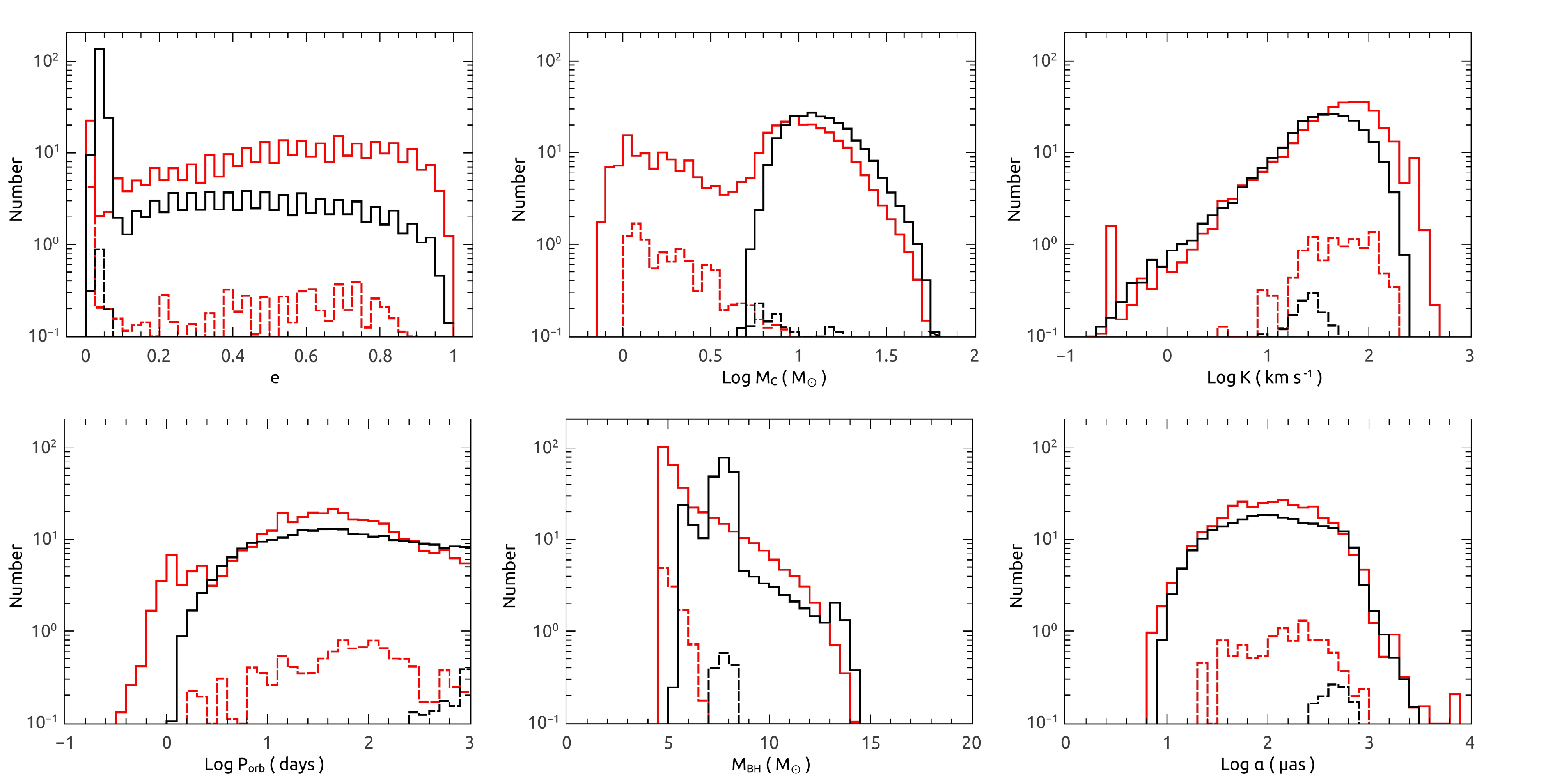}
\caption{Similar to Figure~5, but for detectable BH binaries with optical companions brighter than 20 mag. 
   \label{figure1}}

\end{figure*}

\begin{figure*}[hbtp]
\centering
\includegraphics[width=1.0\textwidth]{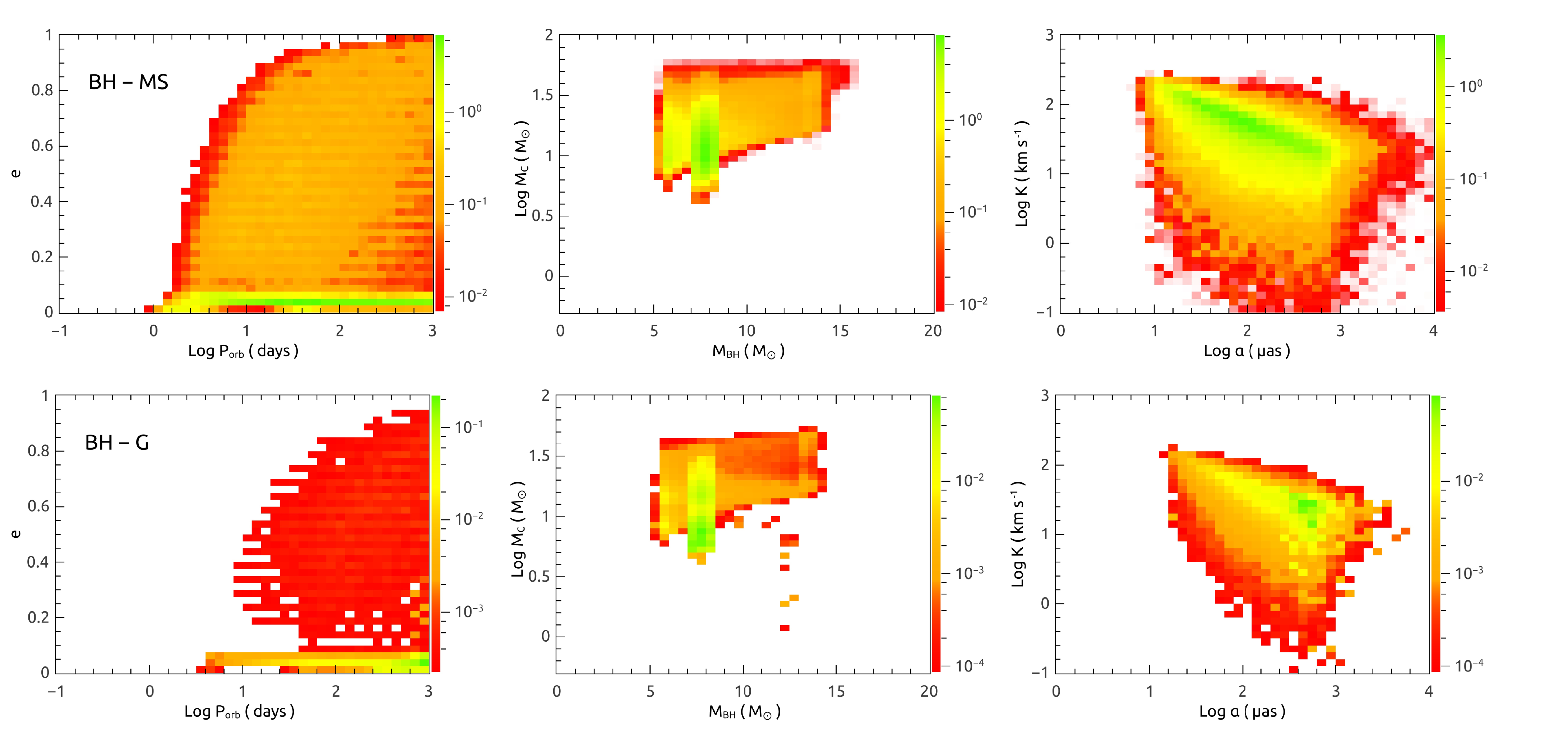}
\caption{Predicted number distributions of detectable detached BH binaries in 
Model A. The left, middle and right panels correspond to the systems distributing in the $ P_{\rm orb}-e $, 
$ M_{\rm BH}-M_{\rm c} $ and $ \alpha - K $  planes, respectively.  The top and bottom panels correspond to 
the systems with MS and giant companions, respectively. The colors in each pixel are scaled according to the number of
detached BH binaries. 
   \label{figure1}}

\end{figure*}

\begin{figure*}[hbtp]
\centering
\includegraphics[width=1.0\textwidth]{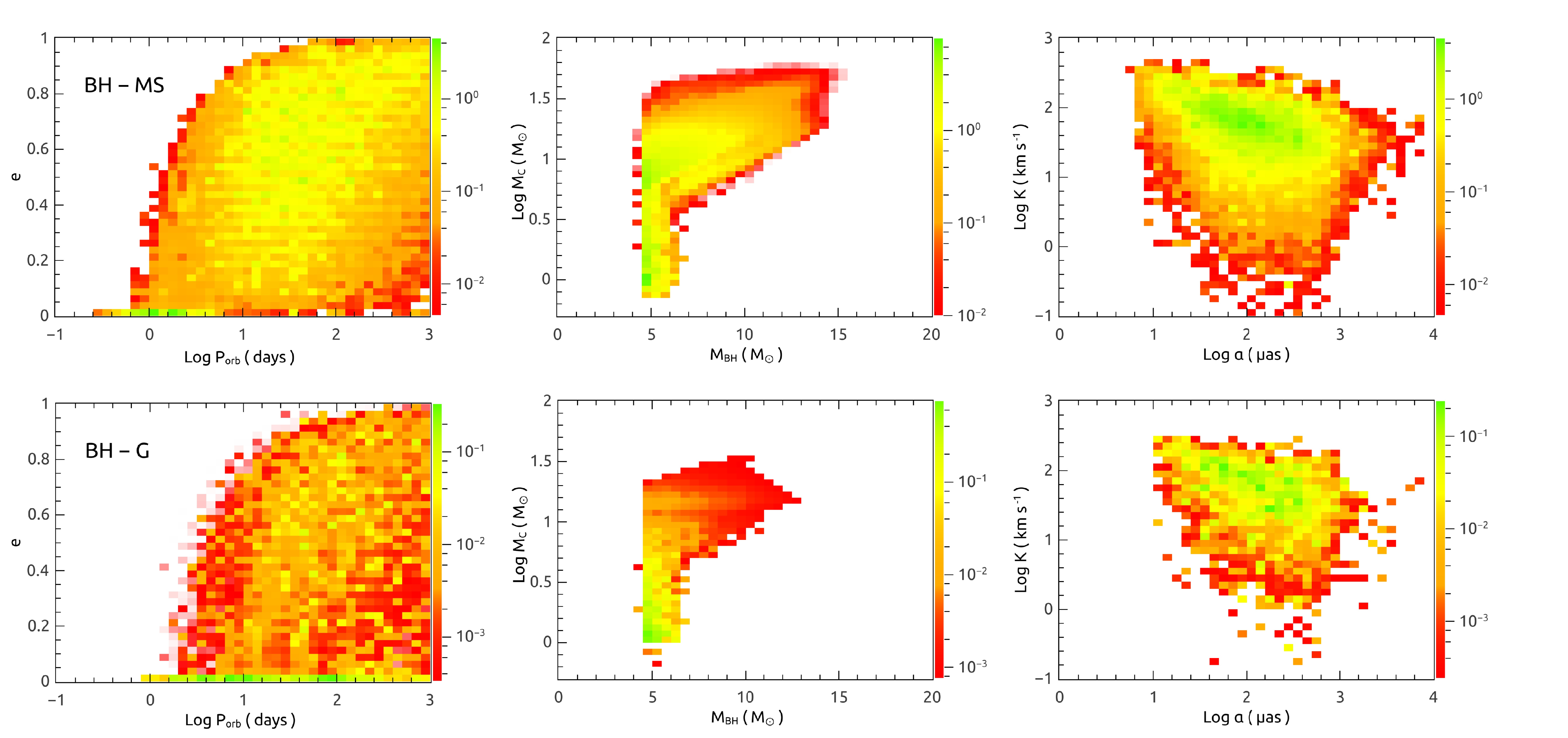}
\caption{Similar to Figure~7, but in Model B. 
   \label{figure1}}

\end{figure*}

In Figure~4 we show the cumulative number distributions of detached BH binaries 
in the Galaxy when gradually 
increasing $ m_{\rm V} $  by an interval of 1 mag. The black and red curves denote the 
results under the assumptions of Models A and B, and the solid and dashed curves correspond to the systems
with MS and giant companions, respectively. We obtain that there are  $ \sim 500 $ detached BH 
binaries in Model A and the number rises to $ \sim 4000 $ in Model B. The
BH systems with MS companions dominate the overall population, only a few percent of them have giant
companions. 

Figure~5 presents the calculated number distributions of all detached BH binaries in the Galaxy,
as a function of the orbital period $ P_{\rm orb} $, eccentricity $ e $, BH mass $ M_{\rm BH} $, 
companion mass $ M_{\rm c} $,  astrometric signature $ \alpha$ and 
radial velocity semi-amplitude $ K $. The black and red curves correspond to Models A and B, and the solid and
dashed curves denote the binaries containing MS and giant companions, respectively. In both models, 
the orbital periods distribute in a wide range of $ \sim1-1000 $ days and quite a number of systems
have nearly circular orbits. In addition, the $ K $ distributions have a peak at $ \sim30-100 \,\rm km\,s^{-1} $
and $ \alpha$ broadly distributes in the range of $ \sim 3- 1000 \,\rm \mu as $. The main differences
between Models A and B are the component masses: the companion masses are $ \gtrsim 5M_{\odot} $ 
and the BH masses cluster $ \sim7-8M_{\odot} $ in Model A, while both the companions and the BHs 
tend to have low masses (peaked at $ \sim1M_{\odot} $ and $ \sim5M_{\odot} $ respectively) in Model B. 

\subsection{Detectable population of detached BH binaries}

Due to the limitation of astronomical instruments, the very dark companions 
in detached BH binaries cannot be detected at present. Based on current high performance of \textit{Gaia}
satellite with a limiting magnitude of 20~mag in the G band, we further discuss the detectable population
of detached BH binaries in the \textit{Gaia} era.  Following \citet{yk18} and \citet{yb18}, we equate the $ Gaia $
band with the V band for the optical companions. For testing the valid of this assumption, we have used the 
color-color transformations of \citet{j10} to compute the \textit{Gaia} G magnitude, and find that the
population size of detached BH binaries is increased by a factor of only a few percent.  
In Figure~6 we show the number distributions of detached BH binaries with optical companions
brighter than 20 mag in the Galaxy, under the assumptions of Models A (black curves) and B (red curves). 
The solid and dashed curves correspond to the systems with MS and giant companions, respectively. 
We can see that a large number of binaries with low-mass
companions are hidden and undetected, the number of detectable binaries drops to several hundreds in 
both models. We emphasize that there are still over 100 detectable detached BH binaries with companion masses 
less than $ \sim5M_{\odot} $ in Model B. 

Figures 7 and 8 present the number distributions of detectable detached BH binaries in the
$ P_{\rm orb}-e $ (left panels), $ M_{\rm BH}-M_{\rm c} $ (middle panels) and $ \alpha - K $  planes 
(right panels) in Models A and B, respectively. The top and bottom panels 
correspond to the systems with MS and giant companions, respectively.
Each panel contains $ 40\times40 $ matrix elements, the colors reflect the number of detached BH binaries 
in the corresponding matrix element by accumulating the product of the birthrates 
with the time durations. Due to tidal interactions, the systems with $ P_{\rm orb} \lesssim 3$ 
days tend to have circular orbits if the companions are still MS stars, while most of binaries can be circularized
if the companions have climbed to the giant branches. For the component masses, Model B particularly predicts
that the low-mass ($ \lesssim 3M_{\odot} $) normal-stars are usually orbited by light ($ \sim4.5-6.5M_{\odot} $)
BHs. There is a tendency that
the larger the $ \alpha $, the longer the $ a $, thus the smaller the $K$.

Recently several groups \citep{ml17,bcl17,yk18,yb18} have explored the prospect of discovering BHs in 
binaries by \textit{Gaia} based on the motions of normal-star companions. 
\citet{ml17} estimated the number of BH binaries that can be detected by \textit{Gaia}  over its five year 
mission to be nearly $ 2\times10^{5} $. However, they did not consider the effects of some important factors
such as interstellar extinction, BH natal kicks and mass transfer process during binary evolution. 
\citet{bcl17} reduced the number to be $ 3800-12000 $ by taking into account detailed treatments 
relevant for the formation of BH binaries, but the effect of 
interstellar extinction was still not incorporated. \citet{yk18} added the effect of
interstellar extinction, and the estimated number of detectable BH binaries significantly 
decreased to $ \sim200-1000 $. Furthermore, \citet{yb18}
showed that their models yield only dozens of detectable BH binaries with luminous companions.

When comparing our obtained population of detectable BH binaries with that by \citet{yk18} 
and \citet{yb18}, we need to point out that there are big differences in the treatment of 
the formation processes of the binary systems. 
(1) Before the formation of the BH binaries, the progenitor systems experience either CE evolution or stable mass 
transfer phases. During the CE phases, both \citet{yk18} and \citet{yb18} adopted a constant 
$ \alpha_{\rm CE} \lambda$ (0.1 or 1.0) for the CE parameter in their BPS calculations. 
The BH systems with low-mass companions could be rarely generated if taking $ \alpha_{\rm CE} \lambda = 0.1$, 
while a fraction of BH binaries could have low-mass companions as the survivors of CE evolution if taking 
an abnormally large value of 1.0 for $ \alpha_{\rm CE} \lambda$, because numerical calculations show that $\lambda$
is usually significantly less than 1 for supergiant stars \citep[e.g.,][]{dt00,prh03,xl10,wva14,wjl16b}.
In our Model~B, the initial masses of 
BH progenitors can be as low as $ \sim15M_{\odot} $, more than 100 BH systems with low-mass 
companions can be created after CE evolution. (2) Both \citet{yk18} and \citet{yb18} indicated
that the companion masses are heavier than $ \sim 8-15M_{\odot} $ due to mass accretion 
if the progenitor systems have experienced
stable mass transfer phases. These are similar to our results in Model A, but we obtain less-massive 
companions with minimal masses of $ \sim 5M_{\odot} $ since 
the rotation-dependent mass transfer mode is adopted during primordial binary evolution. 
Compared to previous works, we further provide the characteristics of BH binaries with giant
companions. Recently the discovery of a candidate BH orbited by a low-mass giant \citep{tk19} seems to
require the formation channel involving a CE phase as proposed in our Model B.

\section{Conclusion}

Based on a BPS method, we have simulated the Galactic population of detached BH binaries with 
normal-star companions. Considering the uncertainties in BH formation physics and relevant
natal kicks, we build four different models to explore the possible effects on the binary population. 
Model A involves the traditional theory of BH formation. In this model the initial masses of BH progenitors are
$ \gtrsim 25 M_{\odot} $, and the primordial binaries with a low-mass secondary cannot survive the spiral-in
phases when they go into CE evolution. The BH progenitor masses can drop as low as 
$ \sim 15M_{\odot} $ in the other models, and this allows the formation of BH binaries with a low-mass companion. 
The predicted number of detached BH binaries in Model A are about one order of magnitude lower than those in 
other models. When only changing the BH natal kicks in Models B, C and D,  
the calculated numbers vary by a factor of less than 3. In addition, there may exist some extra 
factors that can influence the potential number of detached BH binaries. We assume all stars are in 
binaries, while observations show that about $ 70\% $ massive stars are actually in binary 
systems \citep{sd12}, this will slightly reduce the population size. We adopt a constant star formation rate
of $ 3 M_{\odot}\,\rm yr^{-1} $ in the calculations. Many groups \citep[e.g.,][]{sb78,dh06,rw10} obtained the 
star formation rate of the Galaxy varying in the range of  $ \sim 1-5 M_{\odot}\,\rm yr^{-1} $, 
and the actual rate is subject to slight variations along the Galactic age \citep{rp00}. 
These can also change the calculated number by a factor of a few.

Considering that Model A cannot produce detached BH binaries with low-mass 
companions, we summarise our main results from the other models
except Model A as follows. 

1. The birthrates of incipient BH binaries in the Galaxy are in the range of $ 4.5-13\times10^{-5} \rm\,yr^{-1} $
when considering different natal kick distributions for newborn BHs. The systems with companion masses larger 
than $ \sim 5M_{\odot}$ dominate the incipient BH binaries, and the birthrate of 
the binaries with low-mass ($ \lesssim 5M_{\odot} $) companions is of the order 
$ 10^{-6} \rm\,yr^{-1}$. These two groups are roughly separated by the evolution of primordial binaries 
experiencing either stable mass transfer or a CE phase.

2. The overall population of detached BH binaries are dominated by the systems with relatively 
low-mass companions. We 
predict that the total number is over $ 4000 $. If only considering the systems with companions brighter
than 20 mag (i.e., observable by \textit{Gaia}), the number of detached BH binaries reduces to about 
several hundred.

3. Our calculations show that more than 100 detached BH systems with giant companions in the Galaxy and 
among them at least 10 could be detected. Such 
binaries tend to have companions of mass $ \lesssim 5M_{\odot} $ and 
BHs of mass $ 4.5-7M_{\odot} $, whose features are consistent with those of the candidate BH binary
with a giant companion recently discovered by \citet{tk19}.

\acknowledgements
We thank the referee for useful suggestions that helped
improve this paper.
This work was supported by the Natural Science Foundation 
of China (Nos. 11973026, 11603010, 11773015, and 11563003)
and Project U1838201 supported by NSFC and CAS, 
the National Program on Key Research and 
Development Project (Grant No. 2016YFA0400803), and the Natural 
Science Foundation for the Youth of Jiangsu Province 
(No. BK20160611).


\clearpage

\end{document}